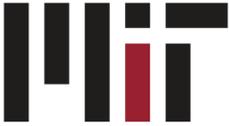

# A Framework for Programmability in Digital Currency

August 1st, 2023

MIT Digital Currency Initiative

# Authors

**Nikhil George**, MIT Media Lab*

**Tadge Dryja**, MIT Media Lab*

**Neha Narula**, MIT Media Lab

*At the time the work was conducted.

# Introduction

As Central Bank Digital Currency (CBDC) research around the world progresses, more countries are exploring use cases for programmability and smart contracts [26,9,1,13]. However, programmable money and programmability in CBDCs are broad terms and there are at least three points of confusion in how to think about, design, and implement programmability: First, programmability is often mistakenly assumed to only be possible in decentralized systems, in conjunction with blockchain technology or distributed ledger technology. Second, programmable money is often conflated with restricted or purpose-bound money – the idea of placing restrictions on how money might be spent by a user, like restricting welfare benefits so they can only be used to purchase items in approved categories [25]. There are other use cases for programmability, like payment-vs-payment and escrow, and restricted money is primarily enabled by permissioning and can be done in the absence of programmability. Third, there is a lack of a systemic framework and vocabulary to discuss different choices in programmability and the tradeoffs and risks of these different choices.

This paper aims to address these concerns by defining programmable money in a more general context, discussing what can be learned about programmable money implementation choices from cryptocurrencies, and providing a framework to discuss programmability choices in other centralized contexts. We offer a definition of programmable money, explain what additional guarantees blockchain technology provides when combined with programmability (though it does not necessarily need to be), and discuss tradeoffs across different locations in the architecture stack where programmability might be implemented: Inside the core system itself, whether (1) via a generic runtime to provide execution of programs supplied by users or clients, also known as smart contracts or (2) by being hard coded directly into the system software. Or, outside the core system software, whether in (3) client-side programs that interact with the system, but do not run inside of it, or (4) with intermediaries that might have more privileged access. We further break down (1) into the types of client-supplied programs supported into various levels depending on the functionality obtained by the offered programmability primitives and discuss the benefits and risks of each level.

# The Rise of Programmable Money

Computers are programmable because one can write carefully formatted instructions which instruct the computer on how to execute a specific task. The computer provides an execution environment that loads inputs and performs the instructions, modifying state and producing results for the caller.[1] Computers today can easily execute programs on databases that store account balances and ledger information for monetary systems. Databases can even embed programmability directly in the form of stored procedures. So what does it mean for a digital asset to be programmable, beyond the type of programmability we already have with generic data records? What we first saw with cryptocurrencies was the unique combination of the following: First, cryptocurrencies have a rich set of instructions (script or a smart contract programming language which compiles to bytecode) in which people can write programs. Second, the monetary asset itself is attached to programs written in that language of instructions, and those conditions govern how the asset is spent. Third, the rules of execution are agreed upon and enforced by a decentralized or distributed network instead of by a single actor or computer, which leads to belief and faith by all parties that the program will execute as specified. Fourth, the instructions are available to use by *anyone* transacting in the monetary asset so that a user can write her own programs governing her asset's transfer. This combination of factors enables a rich ecosystem of applications to be built on top of the transfer of monetary assets that go beyond what we have seen in the traditional financial system.

For example, in Bitcoin, users can write programs to govern the transfer of bitcoins in Bitcoin Script. Instead of merely transferring bitcoin to a new address, they can transfer it into an output encumbered with a script, which specifies conditions that govern the future transfer of the bitcoin. This script will be executed by the Bitcoin network of computers during the process of validating the Bitcoin blockchain. Each Bitcoin node runs a *script interpreter* which evaluates these scripts and is part of the Bitcoin protocol. Because the incentives of the protocol encourage nodes to follow the same ruleset, they will all execute the script as specified by the script interpreter, and achieve the same result. Finally, anyone who holds Bitcoin (and can pay sufficient fees) can create a new transaction spending it to a valid script of their choice, to be executed by the network. Ethereum operates in a similar fashion with the Ethereum Virtual Machine, but with a more expressive instruction set and data model, facilitating the execution of more complex smart contracts that can store arbitrary state and that can be used by many users.

## Defining Programmable Money

Lee [5] describes programmable money as "a unified, coherent product that encapsulates both the storage of digital value and programmability of that value while providing a coherence guarantee."[2] We build on this definition, and our discussion of programmability in cryptocurrencies above, to define programmable money as consisting of four parts:

1. A well-defined format for the digital **storage** of value and data,
2. A well-defined, expressive set of programmable **instructions** for writing programs which access that data and specify the conditions for the movement of that value,
3. A context or environment in which those programs are executed and enforced which provides some **coherence guarantee** that the instructions will execute as specified. In particular, the coherence guarantee should convince participants or a third party that the program will execute as specified,
4. The **permissioning or rules** around who is allowed to create, call and verify the execution of programs.

---

[1] "State" refers to what's stored in the system's memory or on disk, for example in a database. On the nodes running a blockchain this might include account balances, data in smart contracts, or a history of all transactions.

Different programmable money systems may make different choices within the bounds of this definition, for example, on the exact form of data storage or the specific instruction sets provided.

In cryptocurrencies, the coherence guarantee is provided by virtue of the fact that the network is decentralized, and we assume a majority or supermajority of the participants will execute the protocol faithfully, and thus execute the contracts as specified. But a monetary system can provide programmable money even if it is not a decentralized cryptocurrency. The coherence guarantee might be provided by a centralized actor or intermediaries instead of a decentralized network, and it might be governed by regulatory rules instead of technical code (note this might affect people's belief the program will always execute as specified). It might also have some conditions on who is allowed to submit programs.

Some policymakers are interested in programmability because they think it can be used to enforce restrictions on how money can be spent and used [25]. However, these two properties– programmability and use restrictions–are separate. Indeed, one can have a restricted-use non-programmable money (e.g. by adding restrictions to mandated banking APIs), and one can have unrestricted-use programmable money (e.g., by implementing full-featured smart contracts in an openly accessible system). Restrictions attached to users' funds can exist both inside and outside cryptocurrencies and CBDCs, whether they are programmable or not. Further discussion on restricted or purpose-bound money is outside the scope of this paper.[3]

---

[2] Lee defines a coherence guarantee as a mechanism guaranteeing that the technical components of the programmable money product are "inseparable" and that those components are consistently functional, such that the product is stable and coherent for users

[3] Note that applying use-case restrictions on how money can be spent would negatively affect the fungibility, trustworthiness, and stability of the currency and weaken its network effects.

# Programmable money vs. APIs

Today's financial systems sometimes offer Application Program Interfaces, or APIs. In a system with APIs, the system defines some language of requests and promises clients they will give certain responses, depending on how the API is invoked. Note that cryptocurrencies also have services that offer APIs; for example, the most popular API service in the Ethereum ecosystem is Infura, which many applications use to read the state of the blockchain and submit transactions, instead of running their own nodes in the decentralized network, which is more difficult. The line between APIs and a more programmable system is not entirely clear, but there are some differences: The systems behind APIs are often not open source (making it hard to convince users they will always operate correctly), nor are the APIs always available to anyone to use. APIs have limited functionality, and API users have to rely on the API provider to continue to maintain the functionality as promised; for example, an API provider might decide to shut off the service or change how the API operates similar to how Twitter changed APIs for developers in 2012 [18]. APIs also have the downside that they are usually use-case specific, lack composability, and need to be monitored to ensure they actually adhere to their API documentation [23]. In contrast, consider decentralized systems offering smart contracts: They are open source and are permissionless, so not only can anyone use them, but anyone can also join the network to validate that they are executing properly. They can support arbitrary code, and, at least in theory, are difficult to shut down.

The table below compares US bank accounts and cryptocurrencies along the features of programmable money that we defined above.

|  | **Storage** | **Ruleset** | **Environment** | **Permissioning** |
|---|---|---|---|---|
| US Bank accounts | Internal bank database representing accounting of liabilities to users; the data format is not widely known. Users can only access their own data. | Third parties (for example, Plaid) call limited, high-level bank-provided API calls if they exist, or screen scrape.[16] | There is no public verifiability, the bank could reverse or alter execution unilaterally. Users might have recourse via the legal system and regulation via government agencies. | Access to APIs and data is permissioned by the bank. Users provide screen scraping credentials to third parties. |
| Cryptocu-rrencies | Open blockchain of records, which might include unspent outputs, account balances, smart contracts, or other data. Format is well-defined. | More expressive than APIs. A virtual machine or script interpreter executing fine-grained bytecode. The ruleset available is encoded in the blockchain software. | Distributed network consisting of nodes that agree on the ruleset and execute every smart contract. Anyone with a powerful enough computer can run a full node and verify correct execution. Altering execution significantly would require forking the network. | Open to all users, permissioned by transaction fees (individual smart contracts may define their own permissioning) |

*Table 1. Comparing two existing products, US bank accounts and cryptocurrencies across the features of our definition of programmability. US bank accounts might provide APIs but do not provide programmable money.*

Another example of APIs is Open Banking. Money is stored in the form of database entries on computers run by banks, and any programmability offered is done by exposing this database through APIs to specific entities. These APIs are built separately from the database and then are connected through an application running business logic. The UK's Second Payment System Directive (commonly known as Open Banking or PSD2) specifies how regulated third parties can initiate payments directly from customer payment accounts (assuming consent) and access customer data to provide an overview of a customer's payment accounts with different banks in one place. In the UK, the coherence guarantee is provided by a regulated financial institution, who is mandated to provide API access to customer data by the Open Banking Implementation Entity under the purview of the Competitions and Market Authority (CMA).

We argue a centralized system offering programmable money goes beyond existing financial APIs in the following ways. It will:

- a) Provide a high level of commonality and interoperability across programs,
- b) Have a more expressive ruleset that facilitates outcomes that might rely on future, complex changes to state,
- c) Include a coherence guarantee around execution that users can rely on for important financial transactions, and
- d) Provide access to many clients and users.

# Programmable money via client-supplied programs, or smart contracts

Cryptocurrencies showed us a unique way of implementing programmable money – via the on-chain execution of programs, supplied by users, in a virtual environment. Nodes running the blockchain network have execution environments that will execute these programs. For example, Bitcoin exposes Bitcoin script to users, and in Ethereum, users can write more expressive smart contracts which are executed in the Ethereum Virtual Machine (EVM). Depending on the expressivity, it's possible to create new representations of digital assets or tokens *inside* of these programs; for example, in Ethereum, there are smart contracts that create ERC-20 tokens like stablecoins.

As described earlier, it's not clear a "chain" is needed; a centralized system could still support user- or client-provided programs, or smart contracts, without being blockchain-based or even keeping a chain or ledger of transactions (though this might have effects on users' trust of the coherence guarantee). It should, however, have a common execution ruleset which supports a certain set of programmability primitives. The set of primitives defined and available in programs define the interface between the system and the user. We classify these available primitives and the representation of the digital form of money on a linear spectrum from the simplest (Level 1) to the most complex (Level 3).

## Levels of programmability in smart contracts

### Level 1: Signatures tied to fund ownership

Digital assets use *digital signatures* to prove that a user has the ability to move the funds; in its simplest form, the funds are tied to an address derived from a public key, and a user presents the system with a digital signature on a spending transaction, computed using the corresponding private key. Signatures are an authentication mechanism; one could use signatures with a system that only provides APIs. However, Level 1 programmability goes beyond this by tying ownership of funds (representation of money) to authentication via public key cryptography, i.e., the system verifies signatures for each transaction to transfer funds, and knowledge of the corresponding private key is considered ownership of the asset. Level 1 programmability can be enabled with a UTXO-based or account balance system. Using public key cryptography allows users to prove ownership and custody at the end-user level, without needing to involve an intermediary for authentication. If the signature scheme supports it, this enables new programmability use cases such as multi-signature (e.g., both Alice and Bob's signatures are needed to authorize any spending and other more complicated m of n transactions) and limited types of smart contracts, such as those detailed in scriptless scripts [24]. Importantly, note that even the choice of using a specific digital signature scheme might enable minimal programmability.

Zcash shielded transactions, Monero, and Mimblewimble implement Level 1 programmability. All three of these cryptocurrencies aim to preserve transaction privacy and, as a result, do not have scripts or only support minimal scripting.[4]

A Level 1.5 system can be thought of as adding a valid time field to transactions or allowing actions to take place depending on when they are presented to the system. Some use cases that can be enabled with a Level 1.5 system include time-locked transactions, which allow a transaction to be pending and replaceable until an agreed-upon future time, and cross-chain atomic swaps. There is ongoing research being done to enable this functionality in Mimblewimble [2].

**Level 2: Output scripts**

This refers to a system where value is tied to the conditions specified in a script (which might require signatures from keys similar to Level 1).

Bitcoin is an example of a Level 2 system. In systems like Bitcoin, every (non-coinbase) transaction has one or more inputs and outputs, and inputs are previously created transaction outputs. Each output has a small predicate detailing when the output can be spent, called a script. The state of the system is all the unspent funds and spending scripts associated with them (called the set of unspent transaction outputs, or UTXO set). Transactions consume unspent outputs (inputs) and create new unspent outputs once they are confirmed.

Scripts can require signatures from public keys or other conditions to be met, such as time locks. Generally, output scripts cannot reference data outside themselves (e.g., other transactions or balances; one cannot write "send to address B half the amount in address A" without knowing the amounts in each address).

Level 2 systems could support use cases such as simple predicates for spendability (Signature A is valid if Alice's money moves to Bob), escrow, assurance contracts (with oracles; though oracles can be used at all levels), atomic cross-chain swaps, payment channels and Layer 2 networks, derivatives contracts, and other smart contracts with a predetermined set of outcomes.

There have been modifications and extensions to Bitcoin's limited scripting model. For example in Celestia, UTXO scripts can access other data [12], and there are discussions in Bitcoin to expand the expressiveness of output scripts by considering additional script operations that provide *covenants*. A scripting language that supports covenants would allow script authors to restrict the set of scripts that a coin could be further spent to [22]. Some of these techniques enable more complex programs that can be expressed across coin spends. In Bitcoin, transactions are deterministic and usually once valid will continue to be valid regardless of external events (depending only on the existence of the inputs being spent), but additions to the scripting language could change that. Adding in new functionality, like covenants or permitting access to data outside their containing transactions, brings the expressiveness of output scripts closer to Level 3, as described in the next section.

---

[4] The presence of scripts is one of many things that could reduce privacy and distinguish transactions.

**Level 3: Stateful smart contracts**

Level 3 programmability allows for the execution of more expressive client-supplied code and a high degree of introspection, such that the programs executing have access to the full global state of the system. The important features are access to a global state space, the ability to modify things in that state, composability, and pre-specified vs. dynamic allocation of resources.

These provide the most expressivity and fullest set of possible programs and functionality. As an example, a program could query the balances of several other users, compute the average balance, and send funds depending on whether that average meets a threshold. Another example is a decentralized exchange to which users can submit bids and asks, and the contract can execute orders to transfer ownership of tokens when the orders match. Note that the program would not have full custody of the assets being traded and could only transfer assets if orders match, which is different to how centralized cryptocurrency exchanges operate today.

Ethereum is an example of a Level 3 system. The EVM gives developers a model of a shared global computer on which to execute their smart contract code. The global state space is shared and is mutable – programs can edit the state, as opposed to systems like Bitcoin that only offer destroy/create semantics over immutable objects. In addition, smart contracts are composable – one contract can call another one and use its results. But even within Level 3 systems, there is a question of how expressive the language is and what kinds of programs users can write. For example, in Solana, users have to pre-specify the data records their program will touch, while in Ethereum, this can be determined dynamically at runtime. This makes it easier to write smart contracts that can be composed in the future on Ethereum, whereas in Solana smart contracts, authors have to do more up-front work in determining exactly how their smart contracts might be used, and their composability might be limited. However, because of this pre-specification, the Solana nodes know which transactions will conflict and can execute non-conflicting transactions in parallel, getting better performance. It also improves security to limit expressiveness and state access.

**Consideration for policymakers**

Smart contracts might be more useful to users of a CBDC to the extent that it provides a strong coherence guarantee: users can make credible, complex commitments to each other, which will be enforced by the CBDC operator, and programs could be reliable and execute as designed, without intervention by the CBDC operator. Much as users need to trust CBDC operators not to use their control of the system to change monetary balances for public trust, they also should believe the operator will execute programs as submitted. An open question is how much access individual users might have to a CBDC's programmability environment; this question applies to two dimensions. First, whether or not users can read the entirety of the state of the digital currency, including the programs, as they can with public blockchains. This might not be feasible for privacy reasons or if the system is processing many transactions per second and the size of the state is large. Without read access to the state, writing composable programs or getting strong coherence guarantees might be difficult. The second dimension is that there might be some sort of permissioning around who can create or call programs; we discuss this in the intermediated programmability section below.

Stateful smart contracts can be complex with many security considerations, and there are many cases when design flaws, bugs, and other hacks have resulted in a large number of cryptocurrencies being taken against the will and intent of the smart contract authors but using transactions that do adhere to the program code [3,13,11,12,17]. Essentially, the "hacks" obey the letter of the law while violating the spirit [19].

Policymakers and centralized digital currency operators would need to define a dispute resolution process and comprehensive policy framework to determine when and in what capacity an operator might intervene in cases of these hacks. In essence, the entity providing the coherence guarantee is at risk of pressure to reverse that guarantee if things go wrong. In the case of a CBDC, this might put undue pressure on the central bank. Rulemakers for these systems will need to determine if they side with the spirit of the program or the letter of the program. Reliably siding with the spirit of the program and returning funds to those hacked, in the case of smart contracts running on a centralized system, could create a perverse incentive for program designers to focus less on the security of their smart contracts due to the implicit guarantee being provided by the centralized operator. On the other hand, by adhering to the letter of the law and not returning funds in case of hacks, program designers may focus more on program security but may also expend energy lobbying for changes to the rules which govern the operator's stance. The situation with the Parity multi-sig smart contract provides an analog to this scenario where Parity developers stopped all work on the Ethereum ecosystem since Ethereum refused to update the system to recover lost funds [10,5].

While Levels 1 and 2 can present similar considerations to operators, they do so to a lesser extent. With fewer composable and complex smart contracts, there are fewer unexpected results that users may want operators to adjudicate.

# Other locations for programmability

There can be features of programmability in different places in the overall digital currency system. For example, users can write locally-executing scripts to create, sign, and broadcast transactions to a blockchain network. When designing a new digital asset, one can consider providing programmability in different places in the architecture stack. We define three additional locations and explain their benefits and drawbacks: hardcoded within the system rules, with clients, or with intermediaries.

## Within-system programmability

We define *within-system programmability* as the rules embedded within the code of the system itself, separate from any scripting or virtual machine environment for client-facing smart contracts. For example, Bitcoin and Ethereum both have rules for issuance, which is in what quantity and on what schedule new coins should be created; these are *not* subject to user-programming, or even evaluated in the EVM or the Bitcoin script interpreter. Ethereum also has rules on the amount of gas fees that must be paid for a transaction to execute.

In both of these examples, the software running on nodes in the decentralized network enforces these rules; the rules are also known as the *consensus rules* of the system.[5] Users can run validating nodes that verify that the system rules they expect are being applied to the version of the blockchain they see and discard any blockchain which does not follow the expected consensus rules; this is, for example, how the issuance limitation in Bitcoin is enforced. Much of how the blockchain should operate is specified within the consensus rules – including the data format for value (unspent transaction outputs tied to signatures, account balances, etc) and the types of programmable instructions available.

In decentralized systems users enjoy public verifiability, but coordinating a change to the within-system rules in a decentralized system can be challenging. Any user can propose a protocol change with new software, and if a majority of miners (or validators, we use the two interchangeably) decide to run the new software and it is backward-compatible (a soft fork), everyone's software on the network will automatically accept the change, and the system will stay in agreement. If the new software is not backward-compatible (a hard fork), the network will split into two or more chains unless 100% of miners have adopted the new code.

In a centralized currency where end-users do not run validating nodes, they must trust the operator of the system to operate according to the described consensus rules and to not devalue or destroy their data and funds. This is because whoever is running the system, whether it is a single operator or a small group of DLT operators, would have full control over the system's state and consensus rules and could unilaterally make decisions about system operation. This makes it easier to upgrade the system as needed but comes at the cost of public verifiability by users.

---

[5] Note that the consensus rules *also* include the ruleset of the environment for validating and executing scripts or smart contracts.

Both systems need some level of trust but what's different is the party that enforces the coherence guarantee – in the case of a cryptocurrency, it's the decentralized network of users and miners running validating nodes, meaning users can observe the coherence guarantee themselves, while in a centralized CBDC it is the CBDC operator [15]. This makes it harder to provide user enforceability, observability, verifiability, and auditability. One might consider centralized CBDC designs that offer publicly-verifiable proofs of correct program execution to provide a better coherence guarantee, though users would only be able to note if the coherence guarantee was violated, they would not be able to participate and enforce the coherence guarantee directly.

Note that a system with only within-system programmability, though it might be useful, does not meet our definition of offering "programmable money" since users cannot easily create new programs without the operator agreeing to change the rules of the entire system.

**Consideration for policymakers**

A centralized CBDC implementation with a lot of within-system programmability would require a highly trusted and responsible operator or set of operators with strong governance. Examples of within-system programmability in a CBDC could include programming automatic interest rates, programming restrictions on how money can be spent (like restricting spending to only addresses on an allow-list), creating money that has an expiration date, and defining the velocity of currency issuance. However, these are features that affect the flexibility and utility of the money and might affect public trust. The within-system programmability inherent in a CBDC might give the operator too much fine-grained control over the currency people use. Any use of that control would need to be predictable and justifiable to maintain public trust. The question of how much within-system programmability to use is central to how a CBDC fits into the world of money. Credible oversight authorities combined with regulatory and policy frameworks may help set up a basis for a reliable CBDC free from unanticipated within-system changes.

# Client-side programmability

We define client side programmability as the ability to write and execute code on the end-user device or browser. In contrast to smart contracts, this code does not execute in a shared execution environment (though it might call *other* code that does so). An example of doing this in the traditional financial system is the following: If Alice needs to pay Bob recurring monthly payments, she could write up a script and set up a scheduled task on her computer that automatically logs into her bank account on her browser with her username and password and executes a transfer every month; Alice has automated the actions an end-user might take manually on her own browser. Relying on browser interfaces can be unreliable; for example if the bank changes the format of their website, Alice's script might no longer work. The BTCPay Server is another example of this; it's a self-hosted, open-source, non-custodial cryptocurrency payment processor that helps merchants automate invoicing and transaction flows, receiving peer-to-peer payments directly to their wallets.

Client-side programmability can be made more powerful when combined with a shared execution environment for programmability, even if it is very simple. For example, users could use a threshold signature scheme to implement multi-sig in a Level 1 cryptocurrency. The clients can jointly create both a public key and later a signature that the Level 1 chain can validate without needing to know that they reference multiple keys; in fact, the use of threshold signatures is undetectable to the Level 1 chain. Client-side programmability can reduce server load (whether a database or blockchain node); instead of evaluating a longer program, the server just validates what looks like a normal spend transaction, and the complexity is pushed to the participating clients. This also prevents the operator from having visibility into client-side operations. Client-side programmability has the drawback of requiring more complexity on the side of the clients and is not always more efficient. It is also less composable because the program isn't running in a shared state space, and not all functionality can be implemented purely with client-side programmability. One way of thinking about it is that a client-side program is just acting on behalf of its user, and thus cannot act as an arbiter between different parties.

**Consideration for policymakers**

Users should be allowed to automate software running on their own devices. Note that even providing a very simple interface in a digital currency–cryptographic signatures–can provide an opportunity for enhanced client-side programmability. Central bankers and policymakers need to think about how much client-side programmability is possible in a CBDC system and who bears responsibility in case of failure.

# Intermediated programmability

Many central banks are considering intermediated CBDCs, where users access CBDC partially or wholly through an intermediary such as a payment service provider (PSP) or commercial bank. The possible design space around intermediated CBDCs is not yet fully defined, and forms of intermediation surface the question of how to maintain a direct liability on the central bank with intermediation and what additional functionality compared to the current system might be provided [19].

Commercial banks could offer the services of a "programmable bank" that uses the same storage as bank accounts but makes that format, and a rich ruleset of programmable instructions that operates on that data, available to all of its users. Developers could write, and users could run, expressive programs governing the movement of their funds and perhaps even share state to create smart contracts that multiple users can access. An open question is how the commercial bank would provide a coherence guarantee – users might just have to trust the bank to continue to provide this functionality, or policymakers could create rules or regulation. Another question is how to make sure different intermediaries use the same standards, and how to encourage intermediaries to provide cross-organization messaging, in order to prevent walled gardens. Open banking APIs are a step in the direction for a programmable bank, with regulation providing the coherence guarantee. However, in addition to the other downsides of APIs detailed above, open banking APIs are usually very limited in functionality compared to smart contracts. They are also permissioned by a regulated authority that determines who can access them.

## Consideration for policymakers

Much more research needs to be done around programmability and intermediation; however, we list a few of the dimensions that policymakers and researchers can begin to consider:

a. **Who provides the execution environment in such a model, and how is the coherence guarantee enforced?** Some options could include a single shared execution environment for the CBDC (implementing any of the three levels described above) or a model where each bank or PSP has its own separate state and execution space for contract execution. A single shared execution environment would make it easier for smart contracts and other transactions to reference each other, supporting seamless interactions between users using different intermediaries. With separate execution environments, system operators need to consider the level and mechanism of interaction between them. One might imagine a situation like the one we have with different blockchains today, where people create bridges to execute functionality across blockchains; unfortunately these have, in practice, suffered from hacks. Lessons from modular blockchain designs, which are trying to support different execution environments with some shared state, might help think through programmability in this model. Another drawback of intermediary-specific execution environments is relying on the compliance of these institutions to provide the coherence guarantee, which is not always assured.
b. **Who is permissioned to create new programs or smart contracts in such a model?** A CBDC system could require users to have the approval of an intermediary in order to create and submit new smart contracts in a single shared execution environment. Another approach could be that every bank or PSP operating its own environment defines its own user approval process to write programs. These options might hinder innovation since intermediaries could act as gatekeepers.
c. **What authentication or certification do intermediaries provide for users creating and accessing these smart contracts?** If there are transactions and programs that require endorsement to run in the execution environment, the intermediaries could offer endorsement on a per-user, a per-transaction, or a more fine-grained basis.

# Related Work

Many have discussed use cases and potential benefits and risks of programmability in digital payments [26,9,1]. Amazon Web Services and Oliver Wyman [14] adapt the Ethereum technology stack [7] and also discuss how programmability can occur in "different layers of a technology stack" but don't discuss the differences in guarantees across locations and the ruleset of programmable instructions permitted based on system design. OMFIF and Bank of Japan discuss different options for programmability, including differentiating between programmable money, which they define as restrictions on how money might be spent, and programmable payments, which they define as payments with smart contract functionality [13,7]. In this article, we continue to use the original term of "programmable money" to refer to both and discuss programmability in multiple locations. What they call "programmable money" we discuss as restrictions on spending, which might be enforced by the payment system operator in the system software, in which case it falls under our within-system rules category. However, note that in a system with covenants or Level 3 programmability, users (one of whom might be an asset issuer) could program in their own restrictions on how money they issue or hold might be further spent, using smart contracts.

# Conclusion

While some forms of rulesets governing the movement of money have existed for years, it is important for policymakers evaluating CBDCs to think about programmability with a more holistic definition. Policymakers should consider all four dimensions we define when designing such systems, namely the format for the digital storage of value, the set of programmable instructions to specify the conditions for the movement of that value, the coherence guarantee provided by the environment in which those programs are executed and enforced, and the permissioning of who is allowed to create, call and verify the execution of programs. The four potential locations for programmability we lay out highlight the differences in guarantees provided across each of the locations and how policy and regulatory frameworks might be needed where technical guarantees aren't present in the system. While there are many more questions that need to be addressed about programmability in CBDCs, answering these questions and uncovering new challenges will require a coordinated research effort among academia, industry, and the public sector.